# Video-Mediated Emotion Disclosure: Expressions of Fear, Sadness, and Joy by People with Schizophrenia on YouTube

**Liu, Jiaying "Lizzy"**    University of Texas at Austin, USA | jiayingliu@utexas.edu

**Zhang, Yan**    University of Texas at Austin, USA | yanz@utexas.edu

## ABSTRACT

Individuals with schizophrenia frequently experience intense emotions and often turn to vlogging as a medium for emotional expression. While previous research has predominantly focused on text-based disclosure, little is known about how individuals construct narratives around emotions and emotional experiences in video blogs. Our study addresses this gap by analyzing 200 YouTube videos created by individuals with schizophrenia. Drawing on media research and self-presentation theories, we developed a visual analysis framework to disentangle these videos. Our analysis revealed diverse practices of emotion disclosure through both verbal and visual channels, highlighting the dynamic interplay between these modes of expression. We found that the deliberate construction of visual elements—including environmental settings and specific aesthetic choices—appears to foster more supportive and engaged viewer responses. These findings underscore the need for future large-scale quantitative research examining how visual features shape video-mediated communication on social media platforms. Such investigations would inform the development of care-centered video-sharing platforms that better support individuals managing illness experiences.

## INTRODUCTION

Vlogging—video blogging—represents an increasingly prevalent practice for documenting personal stories on video-sharing platforms, offering unique affordances for self-expression through authentic visual presentation and interactive engagement (Naslund et al., 2014; Liu et al., 2023). In health-related contexts, studies have noticed the growing trend of individuals posting health vlogs on platforms like YouTube (Misoch, 2014) and TikTok (Milton et al., 2023), where creators broadcast personal health narratives to chronicle daily challenges, medication experiences, and coping strategies (Huh et al., 2014; Noyes, 2004; Poquet et al., 2018).

Prior studies found that these health vlogs commonly include intense emotional disclosure and span various contexts, including pregnancy loss (Pyle et al., 2021), anxiety (De Choudhury et al., 2014), and depression (Andalibi et al., 2017). Disclosing and sharing emotions through video-sharing platforms enables individuals to visually narrate their stories and create richer, more multidimensional emotional portrayals that facilitate diverse forms of expression (Misoch, 2014) and foster intimate connections between vloggers and viewers (Mickles & Weare, 2020). However, vulnerable populations sharing deeply personal content face significant risks in online spaces, including stigmatization, harassment, and potential exploitation of their emotional disclosures (Competiello et al., 2023). For individuals with mental health conditions, these risks are amplified, as inappropriate responses to their content can exacerbate isolation, reinforce negative self-perceptions, and potentially trigger harmful episodes (Bhuptani et al., 2023).

This tension between the benefits of emotional expression and the risks faced by vulnerable users creates a critical need for research examining how individuals with severe mental health conditions navigate emotion disclosure in digital environments. Our study addresses this gap by investigating the understudied phenomenon of **emotion disclosure through vlogging**—the practice of conveying emotional experiences through video content that integrates verbal expressions with visual elements (Cao et al., 2021). We focus specifically on individuals with schizophrenia who construct and share personal narratives via YouTube vlogs. Schizophrenia manifests through hallucinations, distorted perceptions of reality, and often results in diminished emotional expression and social isolation (Lee et al., 2022; National Institute of Mental Health, 2023). Our findings provide essential insights for designers and platform developers seeking to create safer online environments that support vulnerable users' emotional well-being while fostering genuine community connections. Given that vlogging integrates spoken communication with visual elements, we frame our research questions as follows:

- RQ1: What structural patterns characterize YouTube videos created by vloggers with schizophrenia?

- RQ2: How do vloggers with schizophrenia verbally disclose their emotions in YouTube videos?
- RQ3: How do vloggers with schizophrenia visually disclose their emotions in YouTube videos?

To understand the practices of emotion disclosure through vlogging, we analyzed 200 YouTube vlogs posted by individuals with schizophrenia from 2022-2023. This study makes several contributions to social media studies. (1) We revealed novel insights into video-mediated emotion disclosure, an emerging yet underexplored practice of online disclosure. We identified versatile video narration practices and elucidated how these practices relate to vloggers' emotional states. Our findings can inform future studies to consider emotional contexts in social media research and guide designers in facilitating individuals' nuanced emotion disclosure needs on video-sharing platforms. (2) We discuss the relationships between verbal and visual techniques exhibited in the videos and their potential impacts on viewer interaction, laying groundwork for future research investigating how specific visual elements moderate video-mediated communication. (3) The visual analysis framework provides a foundation for computational research approaches. Future computational studies can apply and expand the framework to examine visual patterns across diverse video content and investigate algorithmic impacts on visibility.

## LITERATURE REVIEW

We reviewed prior studies related to how individuals use vlogging for online care-seeking and then centered on the vlogging practices for emotion disclosure. We then introduced the theoretical lens of self-presentation by Goffman (1959), which we used to investigate the practices of video emotion disclosure of individuals with Schizophrenia.

### Health Vlogging, Online Care-Seeking, and Community Engagement

Video-sharing platforms have emerged as powerful mediators in the care-seeking journey for individuals with health concerns, offering both anonymity and accessibility (Huh et al., 2014). Huh et al. (2014) examined the health vlogging practices of people with chronic illnesses, documenting how individuals candidly share their daily struggles, medication experiences, and coping strategies through video diaries. These authentic visual narratives provide rich, multidimensional portrayals of lived health experiences that text-based forums cannot fully capture.

The multimodal nature of video creates unique opportunities for community building among individuals with health concerns (Misoch, 2014). Videos heighten the salience of interactions by establishing a distinctive social presence and fostering parasocial connections (Lu, 2019), creating immediate, visceral bonds between narrators and viewers (Anjani et al., 2020, Horton & Richard Wohl, 1956, Niu et al., 2021, Short et al., 1976). For individuals with Severe Mental Illness (SMI), these video communities offer crucial social support, validation, and a sense of belonging that may be difficult to access in offline contexts. Through vlogging, individuals with schizophrenia can reclaim their narratives, challenge societal stigmas, and connect with others who share similar experiences.

Thus, this study is motivated by the need to understand how these multimodal elements function in increasingly visual-based social media, particularly how individuals with mental illness utilize visual and verbal techniques to communicate their emotional experiences through vlogging practices.

### Emotion Disclosure through Health Vlogging

Emotion disclosure—the expression of feelings such as anger, sadness, and happiness—plays a pivotal role in the care- seeking process (Chaudoir & Fisher, 2010, Reis & Shaver, 1988). Prior studies around emotion disclosure demonstrate how digital platforms like Twitter, Reddit, and Weibo enable individuals to create supportive communities where users find validation, resources, and connection with others facing similar challenges (Andalibi, 2016, Joinson, 2001, Wang et al., 2011). For individuals with severe mental illnesses (SMI), particularly schizophrenia, emotion disclosure serves an even more crucial function. Schizophrenia manifests through symptoms that dramatically alter an individual's perception of reality and social functioning (National Institute of Mental Health, 2023). Hallucinations and delusions can manifest in disorienting ways, often resulting in reduced emotional expression, social withdrawal, and isolation (Lee et al., 2022). Consequently, individuals living with schizophrenia have an acute need for safe spaces to express their experiences and feelings (National Institute of Mental Health, 2023).

Video emerges as a particularly powerful medium for emotion disclosure due to its intimate and immersive qualities (Liu & Zhang, 2024, Song et al., 2021). Health vlogs frequently capture raw, unfiltered emotions—ranging from moments of triumph to periods of intense anxiety or depression (Liu et al., 2013)—providing viewers with authentic glimpses into lived illness experiences (Woloshyn & Savage, 2020). Berryman and Kavka (2018)'s examination of videos featuring individuals openly expressing vulnerability on camera demonstrates how such visceral emotional displays challenge societal norms and expand the boundaries of acceptable public discourse about mental health.

Through vlogging, individuals with schizophrenia not only exchange valuable insights and coping strategies (Mojtabai & Olfson, 2006) but also share their personal experiences with remarkable candor (Reavley & Jorm, 2011). These digital narratives function as powerful conduits for social support (Sangeorzan et al., 2019), enabling meaningful connections with others who truly understand their unique challenges (Aldao et al., 2010, Cohen and Wills, 1985). In this process, vloggers simultaneously find personal relief and contribute to a broader narrative that challenges deep-seated societal stigmas (Withers et al., 2021), fostering environments where open discussions about mental health can flourish (Livingston & Boyd, 2010).

This study examines how individuals with schizophrenia leverage the visual affordances of vlogging to navigate complex emotional disclosure needs that text-based platforms cannot fully address, providing insights into video-mediated health communication practices that support both personal expression and community connection.

**Emotion Disclosure as Self-Presentation in Health Vlogging**
Visually disclosing emotions and stories about emotions constitutes a sophisticated form of self-presentation. Visual self-presentation involves the strategic use of visual elements to construct and communicate identity within mediated environments. Building on Goffman's dramaturgical framework of self-presentation (Goffman, 1959), this study conceptualizes video platforms as performative spaces where identity is enacted through the deliberate orchestration of multimodal elements (Wan & Lu, 2024). We explore how creators craft and manage impressions through visual aesthetics, performative techniques, and narrative strategies. To systematically analyze visual self-presentation techniques, this study integrates theoretical insights from media research and visual analysis. Drawing on semiotic theory (Barthes, 1968) and color theory, we examine how visual signifiers and chromatic choices shape meaning-making and emotional resonance in mental health vlogs to illuminate how visual elements mediate self-presentation and viewer interpretation.

Semiotic theory provides a framework to understand how visual elements function as signs that convey meaning in mental health vlogs. Research shows that vloggers employ rich visual narration techniques for emotional expression, distinct from previously identified linguistic features in text discourses (Ernala et al., 2017). These visual signifiers manifest in diverse ways across platforms. For example, (Manikonda & De Choudhury, 2017) uncovered distinct imagery patterns that fulfill unique self-disclosure needs, from poignant expressions of emotional distress to raw displays of vulnerability (Andalibi, 2017). The semiotic approach helps explain why certain visual choices resonate with viewers, as studies found that some individuals vocally articulate their emotions while others rely on the silent eloquence of facial expressions and body language (Poria et al., 2017). This theoretical lens also illuminates unique visual communication formats like the "card story" script identified by (Misoch, 2014), where individuals write their mental health struggles on cards held before the camera—a semiotic choice that allows narrators to maintain protective emotional distance while engaging in profound self-disclosure.

Color theory complements semiotics by explaining how chromatic choices influence emotional resonance in mental health vlogs. Beyond the content in the visual representation, color and other visual features are integral as individuals utilize various editing techniques to craft their narratives and expression. The manipulation of color, special effects, and camera angles serves as a powerful medium for self-expression, offering depth and nuance that text alone cannot capture. On platforms like Instagram, (Hong et al., 2020) observed the use of filters and excessive self-presentation techniques, highlighting how visual elements can be leveraged to curate specific personas or emotional states through deliberate color manipulation. This visual language is as diverse as the individuals who employ it, ranging from subtle adjustments to bold artistic choices (Miniukovich & De Angeli, 2014). Color theory explains why even subtle visual cues, such as contrast and hue-related features, can convey significant meaning and emotional states (Ferwerda et al., 2016). These varied chromatic practices underscore the importance of a more nuanced examination of visual techniques in illness narratives and how color choices mediate the relationship between vlogger self-presentation and viewer interpretation.

In this study, we employ an integrated theoretical framework that combines Goffman's dramaturgical perspective with semiotic and color theory to decode how individuals with schizophrenia strategically utilize visual elements as performative tools for emotion disclosure, identity construction, and community engagement through vlogging practices.

**METHOD**
We chose YouTube because it is one of the most trending video-sharing platforms where people with illnesses post vlogs and share their experiences and emotions as suggested by prior studies (Berryman & Kavka, 2018, King & McCashin, 2022). We collected middle-length (4-20 mins) vlogs because such videos include more detailed disclosure of personal stories and intense and intimate narration of emotions (Huh et al., 2014). In comparison, videos such as those introducing medical knowledge about schizophrenia contain little in-depth disclosure.

**Data Collection**

Our search queries combined "schizophrenia" and related terms such as "psychosis" and "schizophrenic" with "vlog," "vlogging," and "story." Using the YouTube Data API v3, we retrieved 555 English-language videos uploaded in 2023. Data collection took place on June 20, 2024. We obtained video transcripts via the YouTube Transcript API1 after downloading the videos using YoutubeDownloader. Following the removal of incomplete data entries (e.g., videos with disabled comments or missing metadata) and institution-created videos, 401 videos remained. We randomly selected 200 videos for analysis in this exploratory study. The average video duration was 9 minutes and 23 seconds.

**Data Analysis**

We employed thematic analysis (Braun & Clarke, 2006), utilizing a systematic approach that combined inductive and deductive elements to examine both visual and narrative content. Our analytical process involved multiple iterative phases (Corbin & Strauss, 2014). We used Excel to organize the codes. Initially, the first author engaged in immersive data familiarization by watching 50% of the video corpus, creating detailed memos that captured both verbal and non-verbal elements. This preliminary phase allowed for identifying emergent patterns while remaining attentive to the multimodal nature of the data (Georgakopoulou & Spilioti, 2016). The coding schema evolved through constant comparative analysis (Glaser & Strauss, 2017). To ensure reliability, our research team, held structured 60-minute analytical triangulation sessions weekly. During these meetings, the first author presented emerging codes where team members could simultaneously view selected video segments. These collaborative discussions facilitated critical reflexivity (Charmaz, 2014). For instance, we discussed and developed codes around patterns in creators' use of lighting techniques during emotional disclosures.

As the codebook stabilized after approximately four weeks of iterative refinement, we moved to axial coding, systematically identifying relationships between conceptual categories and organizing them into higher-order themes. Throughout this process, we actively incorporated theoretical frameworks to guide our analytical approach, for example, drawing from semiotic theory (Barthes, 1968), we examined how meaning was constructed through visual signs within personal spaces in vlogs. Additionally, we interpreted color patterns through established color theory principles, noting correlations between emotional states and visual aesthetics. This theoretically-informed approach enhanced our analytical depth while maintaining grounding in the empirical data.

**Video Structure**. Video structure pertains to the organizational framework and presentation style employed to convey information (Wang et al., 2024a). Our understanding of video structure is informed by prior studies conducting video content analyses (Huh et al., 2014, Niu et al., 2021). For example, Wang et al. (2024b) categorized short videos about intangible cultural heritage according to the videos' meaning construction such as community and innovation. In our context, we focus on how visual and verbal cues are organized to present emotional experiences. Two primary categories, talk-to-camera and in-the-moment, emerged from the analysis. These structural patterns reflect distinct approaches to emotional self-disclosure and narrative construction in mental health vlogs.

**Verbal narrative**. To analyze how vloggers construct emotional narratives through verbal communication, we conducted a thematic analysis (Braun & Clarke, 2006) of video transcripts while referencing the original footage (Corbin & Strauss, 2014). This process involved the identification of expressions and narratives that conveyed emotional states. We noticed two distinct patterns of emotional elaboration: direct emotional expression (explicit statements of feelings, e.g., "I feel overwhelmed") and emotional storytelling (narratives that implicitly convey emotions through personal anecdotes or situational descriptions). To complement our qualitative analysis, we utilized an emotion detection model3 to detect emotions in video transcripts such as joy, sadness, and fear. Based on the computed probabilities of each emotion, we determined the major emotion of each video.

**Visual Frame**. To address RQ3, we sampled one frame every 30 seconds from each video, generating a dataset of 3800 frame images. Our analysis focused on both the visual elements and visual composition within these frames: (1) Inspired by semiotic theory (Barthes, 1968), we analyzed the visual elements within the videos and images by examining how meaning is constructed through visible signs. We systematically coded the visual signs, such as vlogs set in personal spaces and depictions of daily activities, and interpreted the visual in relation to verbal narratives, offering a deeper understanding of how emotion is visually narrated and communicated in vlogs. (2) We also drew upon color theory (Itten, 1961), interpreting how specific color choices contribute to the emotional tone and psychological atmosphere of the videos: for instance, muted colors frequently appeared in depressive episodes. This analysis revealed how content creators deliberately or unconsciously employ color to reflect and reinforce their emotional narratives.

### Ethical Considerations

Although this study qualified as exempt under UT Austin's Institutional Review Board policy, given the sensitive nature of mental health, we protected vloggers' privacy by obscuring facial features and paraphrasing quotes.

### Limitations

One limitation is the potential inaccuracy of automated emotion detection based on video transcripts. As the focus of this study centers on qualitative analysis of video emotion disclosure practices, the impact of this limitation is relatively minor. To mitigate this concern, we consistently referenced the raw videos during data analysis in their entirety to capture vloggers' nuanced expressions through both visual and verbal presentation. Another limitation is that we only analyzed the final product of individuals' information/video creation practices, restricting our ability to understand their complex behavioral and psychological challenges during this process. Future studies could employ methods such as interviews or longitudinal observations to gain deeper insights into vloggers' emotional experiences and content creation strategies throughout their disease healing journey.

## FINDINGS

### Video Structure

We identified two video structures in the schizophrenia vloggers' content: talk-to-camera and in-the-moment. While most vloggers consistently adopt one structure, some transition from the former to the latter over the course of their video creation and posting.

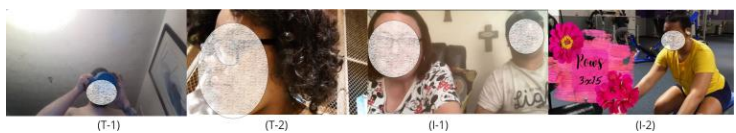

*Talk-to-Camera Structure*

This structure was the predominant format, with 153 videos utilizing this style. As shown in Figure 1 T-1 and T-2, in this approach, vloggers address the camera directly with a static background, creating a confessional atmosphere. These diary-style videos document various aspects of their experiences such as symptom developments, coping strategies, and hospital stays. These video narratives typically follow a structure similar to written diary entries, beginning with temporal and spatial context before delving into recent updates. For instance, one vlogger opened his video with: *"Hello everybody this is Tom, today is September 22, Saturday. It's 11:34 p.m. Eastern Standard Time in Washington and this is my coping with schizophrenia blog entry for today"*, before proceeding to discuss his experiences with hallucinations.

*In-the-Moment Structure*

This structure was found in 47 videos, that visually portray vloggers' lives in a natural and vivid manner as they engage in various activities, such as spending time with family (Figure 1, I-1) or exercising (I-2). This structure creates an accessible and immersive experience for viewers. Notably, nine vloggers directly shared vulnerable and challenging moments on camera, such as crying or experiencing psychosis episodes, which are common struggles for individuals living with schizophrenia. For instance, one vlogger documented her hallucination episode, stating, *"Sharing it feels embarrassing and quite vulnerable because it feels a bit like a personal failure or weakness."* This video garnered over 1.1 million views and 5,147 comments, with many viewers expressing appreciation for her courage.

Relationship between Video Structure and Emotions. We conducted pair-wise Chi-Square tests to check the relationship between video structure and the major emotions conveyed in the video (i.e., fear, sadness, and joy). The results suggested that the proportion of using In-the-Moment structure in the joy group was statistically higher than those in the sadness (p=.0027) and fear group (p=.0003), while there was no statistical difference between the sadness and fear group (p=.7702). We list the proportions in Table 1.

|  | **Fear** | **Sadness** | **Joy** |
| --- | --- | --- | --- |
| **Talk-to-Camera** | 89.6% | 88.1% | 62.5% |
| **In-the-Moment** | 10.4% | 11.9% | 37.5 |

**Table 1. The percentages of video structure used in video groups of different major emotions.**

**Verbal Narrative**

The analysis of the transcript reveals two strategies of verbal emotion disclosure direct emotion expression, explicitly naming and describing emotions, and emotional storytelling, contextualizing emotions through personal experiences. Table 2 listed the emotion words with frequency numbers higher than 10 in the video transcripts and Table 3 included the emerging topics in the stories.

| Emotion | High-frequency words |
|---|---|
| Fear | anxiety (364), crazy (245), worse (171), paranoia (215), fear (116), kill (123), hurt (120), panic (64), violent (64), suicidal (111), hate (55), worry (52), horrible (53), nervous (49), harm (48), mad (45), afraid (80), danger (79), die (71), intense (85), evil (25), death (40) |
| Sadness | ill (578), bad (495), difficult (194), hurt (120), depressed (92), suffering (74), broke (78), struggle (76), terrible (33), witch (40), pain (24), shame (19), nightmare (19), awful (17), distress (15), insane (21), curse (13), brutal (11), abandon (11) |
| Joy | good (967), hope (492), love (435), friend (317), happy (224), enjoy (128), beautiful (134), safe (76), peace (82), excited (59), wonderful (63), helpful (65), content (53), luck (47), success (43), thankful (44), perfect (38), pray (35), comfort (30), laugh (30), proud (29), alive (28), faith (27), inspire (23), powerful (19), accomplish (19), achieve (18), joy (24), encourage (23), happiness (17), embrace (15), magical (15), gratitude (14), pleasant (14), passionate (10) |

**Table 2. Emotion Words Used by Schizophrenia Vloggers**

| Stories | Topics | Example Quotes |
|---|---|---|
| Treatment | Psychiatry & Therapy | "*My psychiatrist suggested therapy to help manage my anxiety.*" |
| | Medications | "*I'm overeating out of anxiety and out of fear. I'm paranoid out of my delusions and my auditory hallucinations.*" |
| Substances Use | Substances Use | "*let's uh one advice I would give everybody, do not even touch any of those substances that are called medicine psychedelics. Just stay away from it.*" |
| Personal Lives | Relationships | "*I feel like I'm a failure as a daughter and she [vlogger's mother] is like why do you feel like that? My parents never got to see me graduate and she's like, why is that? You have to understand that there are standards for a normal person versus somebody dealing with like a debilitating mental illness.*" |
| | Activities | "*I had a long talk with my mom about everything going on.*" |
| | Religions & Death | "*I'm feeling like I'm losing control, overwhelmed by thoughts of death. I've been thinking about shaving my head again. Eventually, the principal came over and told us we could finally go to sleep.*" |
| Recovery Journey | Recovery Journey | "*The voices and the images were greatly lessened, I was doing better in school and was improving every aspect of my life. I got married and I've had got a full-time job my wife and I adopted a kid of foster care... it has not been 23 years since my mission.*" |

**Table 3. Examples of Stories Around Emotion**

Vloggers typically combine both techniques within a single video, though their integration patterns vary. Some open with direct statements like *"Today my anxiety levels are pretty high, my paranoid levels are pretty high, and my depression is pretty severe"* before expanding into narratives, while others begin with stories before explicit emotional disclosure. We observed that certain emotions are intricately linked with the story topics in schizophrenia vlogs. For instance, fear often arises during discussions of psychiatric visits, terrifying symptoms, and hospitalizations. Sadness emerges when vloggers address personal relationships and substance use struggles, while joy surfaces in reflections on recovery and personal growth.

**Visual Frame**

In analyzing how vloggers visually present themselves through video frames, we draw on Erving Goffman's dramaturgical framework (Goffman, 1959) as a theoretical lens. Goffman conceptualized everyday life as a theatrical performance, arguing that social interactions occur on a "front stage" where individuals consciously manage the impressions they make on their audience. This theatrical metaphor proves especially apt for analyzing vlogs (Wan & Lu, 2024), where creators literally perform for their viewers through carefully constructed frames. We decode the frame sampled from the videos (as described in 2.2) by focusing on the three key elements that we identified from the vlogs: vlogger, stage, and style.

|  | **Features** | **Definition** |
|---|---|---|
| **Vlogger** | Anonymity | Whether vloggers reveal their facial identity. |
|  | Demographics | Gender, race, age, etc. |
|  | Identity | Religion, personal beliefs, and cultural background as conveyed in the vlogs. |
| **Stage** | Space | The environment where vlogs take place, including background and setting. |
|  | Activity | The vlogger's actions they engage in during the vlog. |
|  | Other people | The presence of other people and their interactions. |
| **Style** | Color | The use of color schemes and lighting in the frame to convey mood, tone, or emphasis. |
|  | Aesthetics | The overall beauty and visual appeal of the images, related to spatial organization and visual hierarchy within the frame (Metallinos, 2013). |

**Table 4. Visual Analysis of Frames**

*Vlogger.*

The presence of the "self," embodied by the vlogger, is a central and defining element in vlogs, as the vlogger often serves as the main character and focal point of the videos. We found while most of the vloggers disclose their appearance, 19 videos maintain visual anonymity. The examples in Figure 2 showed strategies such as creating dim lighting conditions to blur their appearance (A-1), positioning the camera away from their bodies (A-2), deliberately keeping their faces out of frame (A-3), wearing face masks while recording (A-4), and incorporating anonymous imagery during video editing to conceal their identities (A-5). The major emotions and verbal narratives of these videos are generally similar to other videos, for example, in A-4, the vlogger shared personal experiences to illustrate the importance of having a support network. However, certain cues, such as the voices and outfits, suggest that vloggers maintaining visual anonymity are likely male. Notably, 17 out of the 19 videos analyzed have fewer than 50 views, and 12 videos lack any comments, potentially indicating lower audience engagement of such videos.

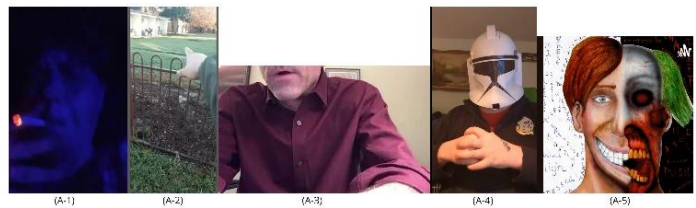

**Figure 2. Examples of Anonymous Framing**

In unanonymous videos, vloggers' demographic and identity markers are visually presented through features such as skin color, age range, and cultural signifiers including religious attire. We observed how these intentionally or unintentionally disclosed identity cues influenced viewer comments. For example, female vloggers often received comments relating to gender-specific stressors; Another vlogger wearing a hijab received comments that sparked conversations about the intersection of faith and mental health, such as *"As a Muslim woman, my faith is my anchor, but it doesn't make me immune to anxiety."* This demonstrates how visible identity markers could shape engagement.

*Stage.*

The stage where vloggers present themselves is reflected in the disclosure of contexts and backgrounds of vlogs, representing a unique affordance of video disclosure compared to text posting on forums. Space reveals where vloggers record the videos, spanning a range of private spaces, such as homes and cars, as well as public areas like gyms and outdoor locations. Vloggers showcase activities, such as art-making, workouts, walks, and jogs. They may also involve other people such as family and friends onto the stage for various purposes.

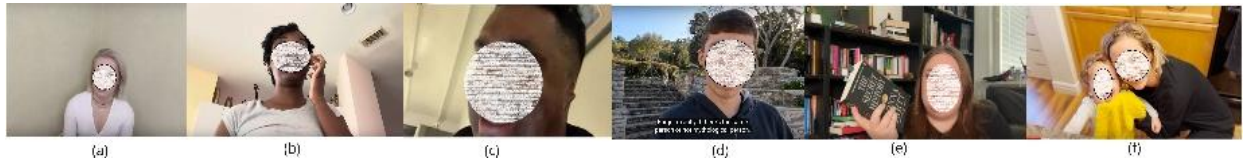

**Figure 3. Examples of Stage**

We observed a spectrum of stage richness as depicted in the frames (Figure 3). In instances where the context is minimally disclosed, vloggers may position themselves in front of a plain background, such as a wall (a) or ceiling (b), or their faces may dominate the frame (c), leaving the surrounding context invisible. Conversely, some vloggers actively highlight their environment by showcasing detailed backgrounds and explicitly discussing their locations (d-f).

As suggested by Goffman, the "stage" as a performance front is intertwined with individuals' self-presentation. We observed that the details of vloggers' stages were closely tied to their verbal narratives. For example, one vlogger (d) filmed videos at his regular walking park while discussing the benefits of being close to nature, while another (e) sat in front of her bookshelf as she recommended books to viewers.

It seems that videos with a carefully crafted stage created opportunities for viewer interaction. Viewers expressed appreciation for vlogger (d)'s immersive walking experiences, with one commenting, *"Wow, what gorgeous scenery! Heavenly! My grandfather had peacocks... I enjoyed walking with you and listening to your talk."* Vlogger (e) even received comments responding to details of her stage that she did not verbally mention, such as, *"I see you have an awesome collection of albums, can you do a video on your album collection?"*

*Style.*

Goffman emphasizes that individuals employ various techniques to shape audience impressions during face-to-face interactions. Similarly, We observed distinct visual styles in the videos, reflected through vloggers' color and aesthetic choices, which may relate to the vloggers' individual stylistic preferences and the emotional resonance of their content.

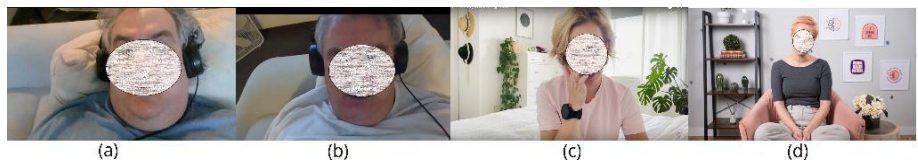

**Figure 4. Examples of Individual Styles**

Colorfulness and brightness seem to resonate with vloggers' narratives about their emotional tone. For instance, low-key lighting, often employed during nighttime recordings, frequently accompanies more somber narratives and discussions of challenging experiences. This is exemplified in Figure 2 A-1, where the vlogger shares their difficult journey with substance use and Schizophrenia.

Individual video creation preferences also seem to influence the vlog styles. Some vloggers, as demonstrated in Figure 4 (c) and (d), maintain a rich, polychromatic palette even during emotionally vulnerable moments, such as crying on camera. In contrast, others opt for more subdued or monochromatic schemes when expressing fear and sadness. One vlogger consistently follows this pattern across their content, as seen in Figure 4 (a) and (b), where they recline on a sofa while sharing their stories. The visual aesthetics of the frames also vary: while frames c-d exhibit a carefully constructed visual syntax, frames a-b present a more spontaneous, less structured approach.

## DISCUSSION

This study explores the emerging practice of emotional disclosure on video-sharing platforms. By focusing on individuals with schizophrenia—a population with significant emotional disclosure needs—we uncovered various strategies they employ to express emotions through vlogging. We identified two primary video structures, talk-to-camera and in- the-moment, which vloggers use to organize their content, as well as two types of verbal narratives: direct emotion expression and storytelling. Additionally, we analyzed frame-level components and techniques, revealing patterns in how vloggers situate their emotion disclosure through configurations of self-presentation, stage setups and stylistic shooting choices. We discuss the implications of these findings for computer-mediated communication (CMC) on video-sharing platforms.

### Supporting Vloggers with Mental Illness for Emotion Disclosure and Self-Presentation

Our research provides novel insights into these behaviors, contributing to the limited literature on video-mediated emotional expression through vlogging among individuals with serious mental illness (SMI). Our findings illuminate how vloggers use videos' richer forms of emotional expression to disclose sensitive stories and feelings around their emotions.

Our analysis revealed emotion-dependent visual and structural behavior patterns. We observed differentiated self-presentation practices across emotional states— particularly, distress-focused videos typically employed a talk-to-camera format with minimal editing compared to videos with neutral emotional content. This selective format usage during emotional distress aligns with Fredrickson's broaden-and-build theory (Fredrickson, 2004), which suggests that negative emotional states constrain cognitive resources, potentially limiting the capacity to produce complex content during these periods. Additionally, we noticed that some vloggers tend to maintain consistent personal visual signatures (specific color schemes, distinctive camera angles, recurring backgrounds) regardless of their emotional state. These visual elements strengthened the creator's identity and fostered viewer connection across emotional fluctuations. For example, one vlogger consistently used warm lighting and close framing during episodes of both distress and joy, creating a recognizable visual identity.

For individuals with SMI, our findings suggest actionable support pathways. Platform designers could develop creator tools specifically designed for emotional disclosure videos, with templates and editing features that facilitate the visual and structural patterns we identified as effective. For instance, YouTube could offer simplified editing templates for distress-state videos that maintain creator visual signatures while minimizing cognitive load. Video-sharing platforms could implement **"emotion-aware" filters** that align with the visual patterns we identified for different emotional states (e.g., subtle color adjustments based on disclosed emotion). Platform designers could develop creator tools for emotional disclosure videos, with templates that facilitate effective visual patterns. Mental health applications could incorporate video journaling features informed by our findings, providing scaffolding through guided prompts and templates tailored to different emotional states—offering simplified options for distress states and more dynamic templates for positive emotional experiences.

### Understanding the Impacts of Visual Features on Visibility and Viewer Engagement on Social Media

This study represents one of the earliest efforts into disentangling the complex interplay of visual elements on video-sharing platforms and their role in shaping viewer engagement. Our observations complement prior research focused on visibility and algorithmic moderation in social media environments. Bishop (2019) suggests that algorithmic systems may prioritize specific aesthetic and emotional presentations, potentially sidelining alternative expressions. Our findings suggest this potential bias might extend beyond textual content to encompass visual elements, possibly influencing which mental health narratives achieve visibility.

Our qualitative analysis indicates possible relationships between visual narration techniques and viewer engagement. Building on prior work, we observed features such as anonymity (Zhang et al., 2021), the sharing of daily routines (Niu et al., 2021), and moments of vulnerability on camera (Huh et al., 2014) that appear to influence perceived authenticity and viewer connection. Additionally, we identified visual elements—such as the physical setup, personal presentation within the frame, and aesthetic choices like color schemes—that may shape emotional disclosure and viewer response. These stylistic choices might influence engagement metrics (e.g., views, likes, and comments), suggesting that visual presentation could play a role in content reception (Wan & Lu, 2024).

Our preliminary observations suggest that videos with certain visual aesthetics—particularly those featuring well-lit environments, vibrant color palettes, and polished editing techniques—might receive different engagement patterns compared to content depicting more raw or unfiltered experiences of mental health challenges. This could potentially create what we might describe as a **"visibility hierarchy"** that may influence which mental health narratives receive more attention. Drawing on visual culture theory (Mirzoeff, 1999), which emphasizes the role of visual representations in shaping perception, identity, and social structures, we speculate that visibility on video platforms may be influenced by algorithmic mechanisms that could interact with mainstream aesthetic values.

These exploratory findings suggest important directions for future Computer-Mediated Communication (CMC) research. Scholars are encouraged to investigate which visual features influence the visibility of mental health narratives on video-based social media platforms through *large-scale computational studies*. Emerging studies demonstrate that LLM-assisted content analysis can effectively extract and annotate visual features from online videos (Liu et al., 2024; Liu et al., 2025). These annotated features can be incorporated into regression models to examine correlations with content visibility and narrative diversity, while appropriate error correction methods ensure reliable statistical testing (TeBlunthuis, 2024). This methodological approach would provide valuable insights into the mechanisms through which platform algorithms shape mental health discourse online.

**Developing Visual Analysis Frameworks for Social Media Video Research**

This study makes methodological contributions to the growing field of social media video research by presenting *a framework for examining video content* with clear dimensions for analyzing framing structure, narrative elements, and visual features. This framework can be applied to future *automated content analysis tools for large-scale video datasets*. Drawing on Hoffman's theoretical framework of self-presentation (Hoffman et al., 2019) and established visual analysis methods in media research (Zuev & Bratchford, 2020), we developed an analytical approach to examine patterns of visual presentation in vlogs. Our framework examines videos across three dimensions: overall framing structure, verbal narratives that shape meaning-making, and detailed visual frame analysis.

Previous research has demonstrated that visual representations significantly influence viewer perception (Hu et al., 2023) and engagement (Li et al., 2024), similar to how linguistic features affect engagement in text-based social media posts. However, the specific visual features that influence community engagement and support, along with their underlying mechanisms, remain underexplored. Existing research has predominantly relied on transcripts, metadata, and comments as proxies for analyzing online videos, while visual content analysis has typically been limited to basic topic identification in images (Bica et al., 2017). Our methodological framework provides a more direct approach to examining visual elements and visual language in vlogs.

While this framework was developed within the context of mental health vlogs, future research can develop and build upon this methodological foundation to analyze various types of video content. Prior HCI studies have examined the complex factors that content creators consider when producing videos, including algorithmic optimization (DeVito, 2022), audience expectations (Barta & Andalibi, 2021), and marketing pressures (Yi & Xian, 2024). Barta and Andalibi's (2024) theoretical lens of visibility illustrating how these considerations influence video content creation. Future research can leverage this framework to examine how these factors manifest in video content and influence the broader dynamics of the social media video economy. Such comparative studies would enhance understanding of visual communication patterns in digital spaces, inform algorithm auditing studies (Karizat et al., 2021), and support platform design decisions that benefit diverse user communities.

**CONCLUSION**

This study examined how individuals with schizophrenia use YouTube vlogs for emotion disclosure, identifying distinct verbal and visual patterns across different emotional states. These multimodal narratives enable more authentic and nuanced emotional expression than text-based media, facilitating viewer connection. Our findings contribute to understanding video-mediated health communication and suggest opportunities for designers to create supportive tools that address the unique disclosure needs of individuals with serious mental illness.

**GENERATIVE AI USE**

We employed Claude for the following purpose(s): polish sentences. We reviewed and revised the returned texts before incorporating them. The authors assume all responsibility for the content of this submission

**ACKNOWLEDGMENTS**

We appreciate the genuine and brave disclosure of people behind the camera, which has greatly inspired the authors and made this project possible. The first author, Jiaying "Lizzy" Liu was supported by the Continuing Fellowship at the University of Texas at Austin and the second author Dr. Yan Zhang was partially supported by the School of Information, John P. Commons Teaching Fellowship.

Livingston, J. D., & Boyd, J. E. (2010). Correlates and consequences of internalized stigma for people living with mental illness: A systematic review and meta-analysis. *Social Science & Medicine*, 71(12):2150–2161.

Lu, Z. (2019). Improving Viewer Engagement and Communication Efficiency within Non-Entertainment Live Streaming. In *The Adjunct Publication of the 32nd Annual ACM Symposium on User Interface Software and Technology*, pages 162–165, New Orleans LA USA. ACM.

Manikonda, L., & De Choudhury, M. (2017). Modeling and Understanding Visual Attributes of Mental Health Disclosures in Social Media. In *Proceedings of the 2017 CHI Conference on Human Factors in Computing Systems*, CHI '17, pages 170–181, New York, NY, USA. Association for Computing Machinery.

Metallinos, N. (2013). *Television aesthetics: Perceptual, cognitive and compositional bases*. Routledge.

Mickles, M. S., & Weare, A. M. (2020). Trying to save the game(r): Understanding the self-disclosure of YouTube subscribers sur- rounding mental health in video-game vlog comments. *Southern Communication Journal*, 85(4):231–243. Publisher: Routledge _eprint: https://doi.org/10.1080/1041794X.2020.1798494.

Milton, A., Ajmani, L., DeVito, M. A., and Chancellor, S. (2023). "I See Me Here": Mental Health Content, Community, and Algorithmic Curation on TikTok. In *Proceedings of the 2023 CHI Conference on Human Factors in Computing Systems*, CHI '23, pages 1–17, New York, NY, USA. Association for Computing Machinery.

Miniukovich, A., & De Angeli, A. (2014). Visual impressions of mobile app interfaces. In *Proceedings of the 8th nordic conference on human-computer interaction: Fun, fast, foundational*, pages 31–40.

Mirzoeff, N. (1999). An introduction to visual culture.

Misoch, S. (2014). Card Stories on YouTube: A New Frame for Online Self-Disclosure. *Media and Communication*, 2(1):2–12.

Mojtabai, R., & Olfson, M. (2006). Treatment Seeking for Depression in Canada and the United States. *Psychiatric Services*, 57(5):631–639. Publisher: American Psychiatric Publishing.

Naslund, J. A., Grande, S. W., Aschbrenner, K. A., and Elwyn, G. (2014). Naturally Occurring Peer Support through Social Media: The Experiences of Individuals with Severe Mental Illness Using YouTube. *PLOS ONE*, 9(10):9.

National Institute of Mental Health (2023). Schizophrenia - National Institute of Mental Health (NIMH).

Niu, S., Bartolome, A., Mai, C., and Ha, N. B. (2021). #StayHome #WithMe: How Do YouTubers Help with COVID-19 Loneliness? In *Proceedings of the 2021 CHI Conference on Human Factors in Computing Systems*, pages 1–15, Yokohama Japan. ACM.

Noyes, A. (2004). Video diary: a method for exploring learning dispositions. *Cambridge Journal of Education*, 34(2):193–209. Number: 2 Publisher: Routledge _eprint: https://doi.org/10.1080/0305764041000170056.

Poquet, O., Lim, L., Mirriahi, N., and Dawson, S. (2018). Video and learning: a systematic review (2007–2017). In *Proceedings of the 8th International Conference on Learning Analytics and Knowledge*, LAK '18, pages 151–160, New York, NY, USA. Association for Computing Machinery.

Poria, S., Cambria, E., Bajpai, R., and Hussain, A. (2017). A review of affective computing: From unimodal analysis to multimodal fusion. *Information Fusion*, 37:98–125.

Pyle, C., Roosevelt, L., Lacombe-Duncan, A., and Andalibi, N. (2021). LGBTQ Persons' Pregnancy Loss Disclosures to Known Ties on Social Media: Disclosure Decisions and Ideal Disclosure Environments.
ASIS&T Annual Meeting 2025    14    Submission Type